\documentclass[aps,prl,twocolumn,groupedaddress,showpacs]{revtex4}
\usepackage{graphicx}

\begin{document}


\title{Quantum Melting of the Charge Density Wave State in 1T-TiSe$_2$}



\author{C. S. Snow, J. F. Karpus, S. L. Cooper$^*$, T. E. Kidd$^+$, and T.-C.
Chiang}

\affiliation{Department of Physics and Frederick Seitz Materials
Research Laboratory, University of Illinois at Urbana-Champaign,
Urbana, Illinois 61801}


\date{\today}

\begin{abstract}
We report a Raman scattering study of low-temperature,
pressure-induced melting of the CDW phase of 1T-TiSe$_2$.  Our
Raman scattering measurements reveal that the collapse of the CDW
state occurs in three stages:  (i) For P$<$5 kbar, the pressure
dependence of the CDW amplitude mode energies and intensities are
indicative of a ``crystalline'' CDW regime; (ii) for 5 $<$ P $<$
25 kbar, there is a decrease in the CDW amplitude mode energies
and intensities with increasing pressure that suggests a regime in
which the CDW softens, and may decouple from the lattice; and
(iii) for P$>$25 kbar, the absence of amplitude modes reveals a
melted CDW regime.
\end{abstract}


\maketitle

There has been a great deal of interest in the relationship
between various diverse and exotic low temperature phases of
strongly correlated systems, including the antiferromagnetic
insulating and unconventional superconducting phases of the high
T$_{\rm c}$ cuprates,\cite{Ginsberg} the charge-ordered insulating
and ferromagnetic metal phases of the manganese
perovskites,\cite{Dagottobook} the orbital-ordered and
ferromagnetic metal phases of the
ruthenates,\cite{Cao2002,Nakatsuji,SnowRuthenate} and the
charge-density-wave (CDW) and superconducting phases of layered
dichalcogenides such as 2H-NbSe$_2$.\cite{Moncton}  Of particular
interest is the exotic phase behavior that is  expected between
fully-ordered (crystalline) and disordered (isotropic) phases as
one tunes the interactions in these systems using some control
parameter other than temperature. These include, for example,
electronically phase-separated regimes,\cite{Dagottobook} and
``quantum liquid crystal'' smectic and nematic phases, which are
expected to be observed between charge-ordered insulating and
`disordered' metallic or superconducting phases as one increases
the interactions between the charge stripes.\cite{Kivelson}
Clearly, therefore, it is of great interest to carefully explore
the manner in which quantum ordered phases collapse, or `melt',
into quantum disordered phases as a function of some control
parameter that tunes the competing interactions in the material at
low temperatures.

In this paper, we report a pressure-dependent low temperature
Raman scattering study of the CDW system 1T-TiSe$_2$, in which we
are able to explore the manner in which the CDW state `melts' with
increasing pressure near T$\sim$0 K.  Because of its layered
structure and simple commensurate CDW phase, 1T-TiSe$_2$ is an
ideal system for such an investigation.  1T-TiSe$_2$ is also of
interest because the CDW transition is not driven by conventional
Fermi surface nesting, but rather by an unconventional mechanism
involving electron-hole coupling and an ``indirect'' Jahn-Teller
effect.\cite{Kidd}  Our low-temperature, pressure-dependent light
scattering approach allows us to explore unique details associated
with quantum mechanical melting of the CDW in 1T-TiSe$_2$. In
particular, this study reveals that the CDW state evolves with
increasing pressure in a manner reminiscent of classical 2D
melting, with `crystalline' and `disordered' regimes, as well as
an intermediate `soft' CDW regime in which the CDW may be
incommensurate with the lattice.  Furthermore, this study
demonstrates the efficacy of pressure-tuned light scattering for
examining a much broader class of quantum phase transitions in
strongly correlated systems.

The 1T-TiSe$_2$ samples used in this study were grown by iodine
vapor transport with a temperature gradient of 570-640 $^{\rm
o}$C.\cite{Kidd}  The sample stoichiometry was verified by x-ray
and resistivity measurements.  The Raman spectra were taken in a
true backscattering geometry with 647.1 nm incident photons.
Variable low temperature, high-pressure measurements were obtained
with a modified SiC-anvil cell inserted into a flow-through helium
cryostat, allowing continuous adjustment of both the temperature
(3.5-300 K) and pressure (0-100 kbar).\cite{SnowRuthenate} Argon
was used as the pressure transmitting medium, and the pressure
inside the cell was determined from the shift of the ruby
fluorescence line; argon is quasi-hydrostatic in the temperature
and pressure range of interest.\cite{Venkateswaran} From the
linewidth variation of the ruby line as a function of pressure and
temperature, we estimate the degree of non-hydrostaticity in the
pressure cell to be 1$\%$ or less.

\begin{figure}[tb]
\centerline{\includegraphics[width=8cm]{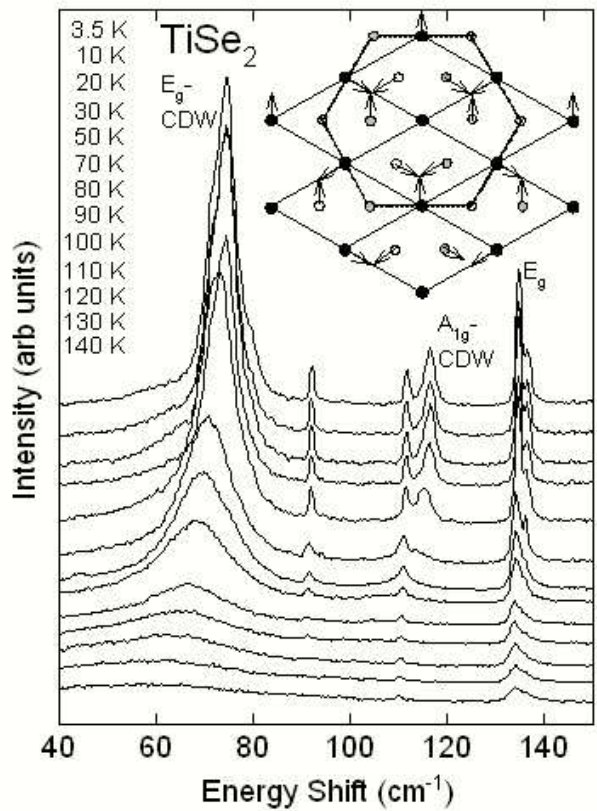}}
\caption{Temperature dependent Raman scattering spectra of
TiSe$_2$.  The inset shows the displacement pattern associated
with the CDW distortion, for Ti atoms (solid), Se atoms above a Ti
layer (open), and Se atoms below a Ti layer (shaded).}
\label{tise2tempraman}
\end{figure}
\begin{figure}[tb]
\centerline{\includegraphics[width=6cm]{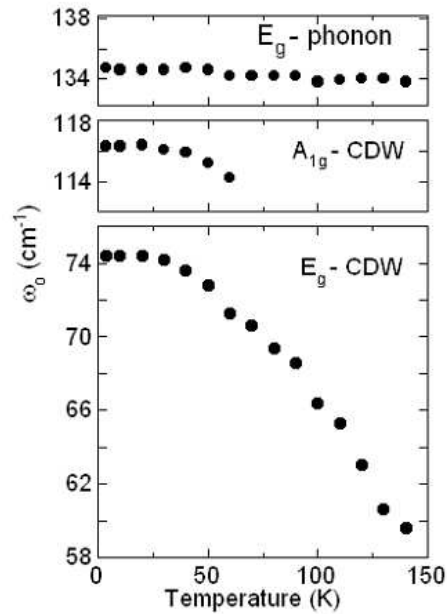}}
\caption{Temperature dependence of the peak energies of the (top)
$E_{g}$ optical mode, (middle) $A_{1g}$ CDW-amplitude mode, and
the (bottom) $E_g$ CDW-amplitude mode.} \label{tise2tempsumm}
\end{figure}

Figure~\ref{tise2tempraman} shows the temperature-dependent Raman
scattering spectrum below the CDW transition temperature at
T$_{\rm C}$ $\sim$ 200 K.  Several new modes develop in the CDW
phase.  Of particular interest are an E$_g$ mode near 75 cm$^{-1}$
and an A$_{1g}$ mode near 115 cm$^{-1}$. These two modes are
CDW-coupled ``amplitude'' modes associated with the zone-boundary
transverse acoustic phonons from the $L$-point in the Brillouin
zone, which are folded to the zone-center due to the formation of
the CDW superlattice.\cite{Holytise2,Sugai} That these two modes
are indeed coupled to the CDW mode is confirmed by their
temperature dependence in Fig.~\ref{tise2tempsumm}: Both the 75
cm$^{-1}$ E$_g$ and 115 cm$^{-1}$ A$_{1g}$ amplitude mode energies
soften as the temperature is increased toward the CDW transition
temperature. By contrast, the E$_g$ zone-center optical phonon
near 134 cm$^{-1}$ is nearly temperature-independent, indicating
that it is not strongly influenced by the development of the CDW
state.

A few comments concerning the two CDW amplitude modes observed in
Fig.~\ref{tise2tempraman} will help in the interpretation of the
pressure-dependent light scattering data.  First, the CDW
amplitude modes are excitations of the CDW ground state, involving
fluctuations of the CDW state that modulate the amplitude of the
charge density wave. In particular, the A$_{1g}$ amplitude mode
near 115 cm$^{-1}$ is a breathing mode distortion, and therefore
involves fluctuations of the CDW amplitude that preserve the
symmetry of the CDW ground state.  Consequently, the A$_{1g}$
amplitude mode displacement pattern is the same as the static CDW
displacement pattern shown in the inset of
Fig.~\ref{tise2tempraman}.  On the other hand, the E$_g$ amplitude
mode involves out-of-phase fluctuations of the CDW amplitude away
from the ground state symmetry.  In combination, these two
amplitude modes serve as ideal ``probes'' with which to study the
stability and elasticity of the CDW state as it evolves, and
eventually melts, as a function of increasing pressure.

\begin{figure}[tb]
\centerline{\includegraphics[width=6cm]{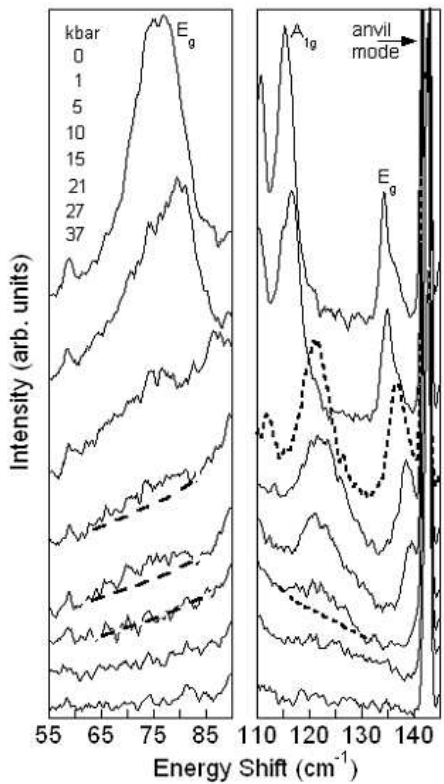}}
\caption{Pressure-dependence of the Raman spectra of TiSe$_2$ at
T= 3.5K.  The intense mode near 143 cm$^{-1}$ is an anvil phonon
mode. The long-dash lines depict the approximate background levels
under the amplitude modes shown.} \label{tise2pressraman}
\end{figure}
\begin{figure}[tb]
\centerline{\includegraphics[width=8cm]{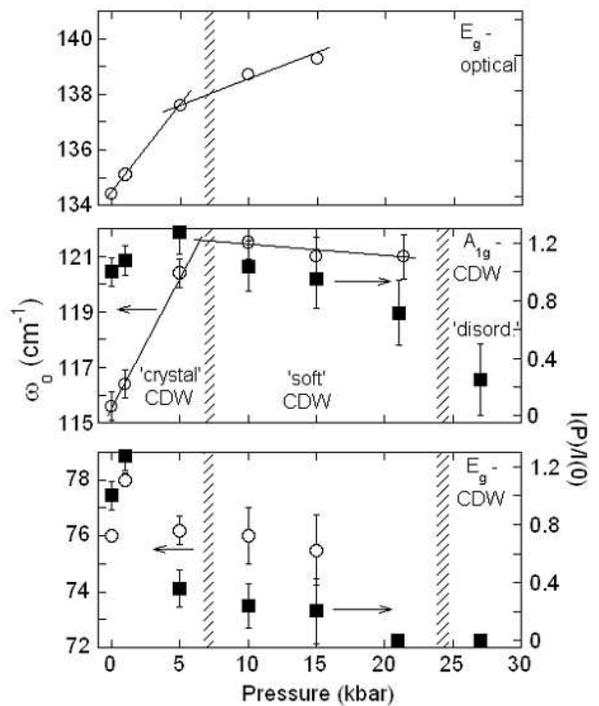}}
\caption{Pressure-dependence of the peak energies (circles) and
normalized integrated intensities, I(P)/I(P=0) (squares) of the
E$_g$ optical mode (top), and A$_{1g}$ CDW-amplitude mode
(middle), and E$_g$ CDW-amplitude mode (bottom) at T=3.5K.}
\label{tise2press_summ}
\end{figure}

Figure~\ref{tise2pressraman} shows the pressure dependent Raman
spectra of TiSe$_2$ at 3.5 K.  The energy and intensity of the 115
cm$^{-1}$ A$_{1g}$ and 75 cm$^{-1}$ E$_g$ CDW amplitude modes,
along with the 134 cm$^{-1}$ E$_g$ optical phonon mode, are
summarized in Fig.~\ref{tise2press_summ}. One of the chief effects
of pressure on the low temperature Raman spectrum of 1T-TiSe$_2$
is the gradual suppression of the 75 cm$^{-1}$ and 115 cm$^{-1}$
CDW amplitude mode intensities with increasing pressure, and the
complete collapse of the CDW state near a T$\sim$0 critical
pressure of P$^*\sim$25 kbar.  This value of the critical pressure
is similar to that observed in pressure-induced CDW-to-metal
transitions measured using transport measurements, including
NbSe$_3$ (P$^* \sim$24 kbar)\cite{Yasuzuka} and
Lu$_5$Ir$_4$Si$_{10}$ (P$^* \sim$21 kbar).\cite{Lue}

Even more interesting than the observed pressure-induced collapse
of the CDW phase, is the manner in which this collapse occurs.
This process can be carefully studied by examining the energies
and integrated intensities of the 75 cm$^{-1}$ E$_{g}$ and 115
cm$^{-1}$ A$_{1g}$ amplitude modes as a function of pressure,
which are summarized in Fig.~\ref{tise2press_summ}(b) and (c),
respectively.  For comparison, the pressure-dependent energy of
the 134 cm$^{-1}$ optical phonon mode is illustrated in
Fig.~\ref{tise2press_summ}(a).

Notably, Fig.~\ref{tise2press_summ} reveals several regimes of
behavior associated with the pressure-induced (T$\sim$0) quantum
melting of the CDW state in 1T-TiSe$_2$: (i) {\it Crystalline CDW
regime} - From P=0 to 5 kbar, the A$_{1g}$ CDW amplitude mode's
intensity increases, and its energy increases at a rate of
approximately $d \omega_o/dP \sim$ + 1 cm$^{-1}$/kbar. This
behavior is consistent with increased stiffening of the CDW state
with increased pressure, characteristic of a typical
``crystalline'' phase. Significantly, the rate at which the
A$_{1g}$ amplitude mode energy increases with pressure is similar
to that observed for the 134 cm$^{-1}$ optical phonon mode
(Fig.~\ref{tise2press_summ}(a)), providing evidence that the CDW
remains commensurate with the lattice in this regime. (ii) {\it
Soft CDW regime} - Between P = 5 to 25 kbar, the A$_{1g}$
amplitude mode exhibits a number of interesting and anomalous
changes as a function of increasing pressure: its energy softens -
revealing an anomalous, slightly negative Gr\"uneisen mode
parameter d$\omega_o$/d$P$ in this regime - its intensity
decreases rapidly, and its linewidth increases substantially.  All
of this behavior is consistent with a distinct softening of the
CDW state to breathing-mode fluctuations of the CDW amplitude in
this regime. Equally interesting is the fact that the energies of
the A$_{1g}$ amplitude and E$_{g}$ optical phonon modes exhibit
distinctly different pressure-dependencies in this regime,
suggesting that the CDW decouples from the lattice (i.e., becomes
incommensurate) in this regime.\cite{commensurate}  Evidence for a
pressure-induced transition to an incommensurate CDW phase has
also been observed in 2H-TaSe$_2$ by high-pressure x-ray
scattering.\cite{McWhan}  Interestingly, the E$_g$ amplitude mode
also exhibits a particularly anomalous pressure-dependence in this
regime, including a decrease in energy and a rapid reduction in
intensity with increasing pressure.  The rapid disappearance of
the E$_g$ mode, in particular, indicates that out-of-phase
fluctuations of the CDW amplitude are not well-defined excitations
above roughly P$\sim$5 kbar, even though there is clearly some
vestige of the CDW state at these pressures, as evidenced by the
presence of the A$_{1g}$ amplitude mode.  This may indicate a
breakdown of long-range CDW order in this phase regime; indeed,
calculations of 1D CDW systems show that a softening of the CDW to
shear deformations, and a consequent breakdown of long-range
translational order, occurs when the coupling between CDW stripes
reaches a critical value.\cite{ZaitsevZotov}  In sum, the observed
pressure-dependencies of the CDW and lattice modes are
characteristic of a regime in which the CDW has not only begun to
soften, or `melt', but in which the CDW may have an increased
propensity to move (`slide') in response to an applied field due
to decreased pinning to the lattice. (iii) {\it Disordered regime}
- Finally, above P$^*\sim$ 25 kbar, there is no evidence in the
spectra for long- or short-range CDW order, suggesting that the
CDW state has melted completely into a disordered phase.

It is interesting to compare the pressure-induced `melting'
process described above for 1T-TiSe$_2$ to melting in other
systems.\cite{HalperinNelson,Dai} For example, calculations of
classical melting in 2D solids suggest the presence of a
``hexatic'' phase - in which long-range orientational order is
preserved, but long-range translational order is lost -
intermediate between the crystalline and disordered phases.  This
topological phase arises because dislocations cause translational
order to decay exponentially, but cause a much weaker suppression
of orientational order. Such a hexatic-like phase has indeed been
observed as a function of increasing disorder ($x$) in the
Nb$_x$Ta$_{1-x}$S$_2$ system using scanning-tunnelling-microscopy
(STM):  these measurements reveal a system that is, in turn,
crystalline (0 $<x<$ 0.04), hexatic (0.04 $\leq x \leq$ 0.07), and
amorphous ($x>$ 0.07), as a function of increasing disorder ($x$).

While the above phase behavior in 2D systems is reminiscent of the
phase regimes we observe as a function of pressure in 1T-TiSe$_2$,
two important points should be made with respect to this
comparison. First, in contrast to the examples above, the melting
process we observe in 1T-TiSe$_2$ is driven at T$\sim$0 K by
pressure-tuning the competing interactions in this system.  To
understand the nature of this competition, note first that the
zero-pressure CDW state in 1T-TiSe$_2$ is unconventional, arising
from an indirect Jahn-Teller-type interaction that splits and
lowers the unoccupied conduction band:\cite{Kidd} as a result of
the electron-hole interaction  between the conduction and valence
bands, the lowering of the split conduction band ``repulses'' and
flattens the valence band, resulting in a lowering of the system's
energy, and the formation of a small gap CDW state.  Upon applying
pressure to this CDW state, one expects several regimes of
behavior:  at low applied pressures, increasing pressure will
increase the matrix element associated with the Jahn-Teller
interaction; this will result in a further lowering the conduction
band, and via the electron-hole coupling, a lowering of the
valence band and a consequent stiffening of the CDW state.  This
behavior is similar to that observed in the ``crystalline CDW''
regime of Fig.~\ref{tise2press_summ}.  As the pressure is
increased beyond some critical pressure, however, the increasing
strength of the Jahn-Teller interaction will eventually overwhelm
the electron-hole interaction between the conduction and valence
bands, leading to a collapse of the CDW gap, and a
pressure-induced transition to a metallic phase in which the CDW
distortion is completely suppressed. Most significantly, our
results indicate that prior to the complete collapse of the CDW
gap above P$^*$, there is a distinct ``soft CDW'' phase regime in
which the CDW loses its stiffness; this is likely to result from
an increase in fluctuations of the CDW near P$^*$.  In theoretical
support of this, Zaitsev-Zotov {\it et al}. have shown that
increased coupling between CDW 'stripes' leads to an increase in
dynamic fluctuations and a decreased CDW
stiffness.\cite{ZaitsevZotov} Interestingly, such long wavelength
lattice fluctuations are expected to destroy long-range
translational order but preserve long-range orientational
order.\cite{Mermin}

Second, the degree of interlayer coupling in 1T-TiSe$_2$ should
influence the nature of the quantum melting
transition.\cite{mcmillan} For example, while classical melting of
a 2D system occurs via a continuous phase transition with an
intermediate topological phase regime, classical 3D melting occurs
via an abrupt first-order phase transition. Significantly, the
continuous nature of the pressure-induced evolution of the CDW
state we observe in 1T-TiSe$_2$, and evidence for an intermediate
``soft CDW'' regime in which there may only be short-range CDW
order, suggests that pressure-induced quantum melting of the CDW
in 1T-TiSe$_2$ occurs in a manner akin to classical 2D melting.

In summary, Raman scattering studies of T$\sim$0 pressure-induced
melting of the CDW state in 1T-TiSe$_2$ reveals both a
low-pressure (P$<$5 kbar) ``crystalline'' regime, and a
high-pressure (P$>$25 kbar) ``disordered'' regime.  Most
interesting, however, is evidence for an intermediate pressure
regime (5$\leq$P$\leq$25 kbar) in which the CDW loses its
stiffness and may have no long-range translational order.

We thank M.V. Klein and E. Fradkin for useful discussions.  This
work was supported by the National Science Foundation under grant
DMR02-44502, and by the Department of Energy under grant
DEFG02-91ER45439.

\noindent$^*$ Corresponding author: s\_cooper@mrl.uiuc.edu \\$^+$
Present address:  Brookhaven National Laboratory
\vspace{-0.7cm}
\bibliography{basename of .bib file}

\begin{thebibliography}{}
\vspace{-0.7cm}
\bibitem{Ginsberg} {\it Physical Properties of High-Temperature Superconductors I-V}, ed. D. M.
Ginsberg, World-Scientific, Singapore (1989).

\bibitem{Dagottobook}  E. Dagotto, {\it Nanoscale Phase Separation and Colossal Magnetoresistance: The physics
  of Manganites and Related Compounds} Springer, Berlin-Heidelburg (2003).

\bibitem{Cao2002} G. Cao {\it et. al.}, Phys. Rev. B {\bf 67},
060406 (2003).

\bibitem{Nakatsuji} S. Nakatsuji and Y. Maeno, Phys. Rev. Lett. {\bf 84}, 2666 (2000).

\bibitem{SnowRuthenate} C. S. Snow {\it et. al.}, Phys. Rev. Lett.
{\bf 89}, 226401 (2002).

\bibitem{Moncton} D. E. Moncton, J. D. Axe, and F. J. DiSalvo, Phys. Rev. Lett. {\bf 34}, 734
(1975).

\bibitem{Kivelson} S. A. Kivelson, E. Fradkin, and V. J. Emery,
Nature {\bf 393}, 550 (1998).

\bibitem{Kidd}  T. E. Kidd {\it et al.}, Phys. Rev. Lett. {\bf
88}, 226402 (2002).

\bibitem{Venkateswaran} U. Venkateswaran {\it et. al.}, Phys. Rev. B. {\bf
33}, 8416 (1986).

\bibitem{Holytise2}  J. A. Holy {\it et. al.}, Phys. Rev. B {\bf
16}, 3628 (1977).

\bibitem{Sugai} S. Sugai {\it et. al.},  Sol. St. Commun. {\bf 35}, 433
(1980).

\bibitem{Yasuzuka} S. Yasuzuka {\it et. al.}, J. Phys. Soc. Japan {\bf 69}, 3470
(2000).

\bibitem{Lue}  C. S. Lue {\it et. al.}, Physica C {\bf 364}, 243
(2001).

\bibitem{commensurate}  An incommensurate CDW state should result in
additional ``phase'' modes in the spectrum, although such modes
will likely be too weak to observe due to the experimental
conditions and the intensity of the CDW modes in this pressure
regime.

\bibitem{ZaitsevZotov} S. V. Zaitsev-Zotov {\it et. al.}, Phys. Low-Dim. Struct. {\bf 1-2}, 79
(2002).

\bibitem{McWhan} D. B. McWhan {\it et al.}, Phys. Rev. Lett. {\bf 45},
269 (1980).

\bibitem{HalperinNelson}  B. I. Halperin and D. R. Nelson, Phys. Rev. Lett. {\bf 41}, 121
(1978).

\bibitem{Dai} H. Dai, H. Chen, and C. M. Lieber, Phys. Rev. Lett
{\bf 66}, 3183 (1991).

\bibitem{Mermin}  N. D. Mermin, Phys. Rev. {\bf 176}, 250 (1968).

\bibitem{mcmillan}  W. L. McMillan, Phys. Rev. B {\bf 12}, 1187
(1975).
\end{thebibliography}

\end{document}